%% file: VVatCMS_dpf2013.tex
\documentclass[12pt]{article}
\usepackage{graphicx}

\newcommand\pubnumber{DPF2013-143}
\newcommand\pubdate{\today}

\def\tamu{ Department of Physics and Astronomy\\
Texas A\&M University, College Station, TX, USA}
\def\collaboration{\footnote{On behalf of the CMS Collaboration}}

\def\Title#1{\begin{center} {\Large #1 } \end{center}}
\def\Author#1{\begin{center}{ \sc #1} \end{center}}
\def\Address#1{\begin{center}{ \it #1} \end{center}}

\newcommand\pubblock{\rightline{\begin{tabular}{l} \pubnumber\\
         \pubdate  \end{tabular}}}
\newenvironment{Abstract}{\begin{quotation}  }{\end{quotation}}
\newenvironment{Presented}{\begin{quotation} \begin{center} 
             PRESENTED AT\end{center}\bigskip 
      \begin{center}\begin{large}}{\end{large}\end{center} \end{quotation}}
\def\Acknowledgments{\bigskip  \bigskip \begin{center} \begin{large}
             \bf ACKNOWLEDGMENTS \end{large}\end{center}}

\input econfmacros.tex

\begin{document}
\begin{titlepage}
\pubblock

\vfill
\Title{Measurements of VV Boson Production And Self-Interactions in The Semileptonic Channel at CMS}
\vfill
\Author{ Osipenkov, Ilya \collaboration}
\Address{\tamu}
\vfill
\begin{Abstract}
We present a summary of the latest measurements of the VV cross-sections, where V$=$W or Z and one of the bosons decays into a pair of jets while the other decays leptonically. The data sample(s) correspond to proton-proton collision events collected with the CMS detector at $\sqrt{s} = 7$~TeV. Subsequent searches for Anomalous Triple Gauge couplings, which allow us to probe the non-Abelian structure in the Electroweak Sector, are described. We present the exclusion limits on the corresponding couplings.
\end{Abstract}
\vfill
\begin{Presented}
DPF 2013\\
The Meeting of the American Physical Society\\
Division of Particles and Fields\\
Santa Cruz, California, August 13--17, 2013\\
\end{Presented}
\vfill
\end{titlepage}
\def\thefootnote{\fnsymbol{footnote}}
\setcounter{footnote}{0}

\section{Introduction}
Triple gauge boson couplings, which determine the self-interactions
of W and Z bosons, are fixed by the gauge symmetry of the standard model (SM). Consequently, pair production of vector gauge bosons
allows a direct test of the electroweak sector of the SM~\cite{PhysRevD.48.2182}. Observation
of anomalous triple gauge boson couplings (aTGCs) would correspond to the presence of physics beyond the SM.

In this paper we summarize the first measurement of WW+WZ diboson
production in pp collisions in the semileptonic final state at
the Large Hadron Collider (LHC)~\cite{OurArticle}. Here
one W boson decays leptonically while the other boson
(W or Z) decays hadronically ($jj$), giving rise to two energetic jets in
the final state. The measurement is particularly challenging due to the fact that vast majority of events
in this final state originate from the $W+Jets$ irreducible background. The production is dominated by quark-gluon
scattering (vs. $q\bar{q}$ at the Tevatron), which further increases its relative contribution. The predicted cross-sections
are $\sim 3\times 10^{4}$ pb for $W+Jets$ vs. $\sim 70$ pb for VV at NLO. Overall, the Signal to Background ratio is much worse at
the LHC and stronger cuts as well as improved analysis techniques are needed.

The advantage for studying the diboson production in the semileptonic final state,
over the purely leptonic final ones, is the larger
branching fraction of W and Z bosons to quarks. The process also permits a direct measurement of the boson transverse
momentum ($p_{T}$). Tha analysis can be expanded to include boosted topology, complimentary final states ({\it e.g.} $W(\to jj)+\gamma$, $V(\to jj)+Z(\to MET)$) as well as Vector Boson Fusion studies at $8$ and $13$~TeV. The current goals are to confirm the WV signal in the semileptonic events at CMS and set limits on corresponding aTGCs.

\section{Object Selection And Backgrounds}

The data were collected with a series of single-lepton triggers mostly
using $p_T$ thresholds of 24~GeV for muons
and 25--32~GeV for electrons. In order to select events with on-shell W bosons we require lepton $p_{T}>$25~GeV (35~GeV), $\eta<$2.1 (2.4), $W$ transverse mass $>$30~GeV (50~GeV), MET$>$25~GeV (30~GeV) and secondary lepton veto in the muon (electron) channel. Exactly two AK5 jets are reconstructed subject to PileUp corrections, isolation from leptons $>$0.3, $p_{T}>$35~GeV, $\eta<$2.6 and jet b-tag veto on the secondary vertex. We implement additional quality cuts in order to to enhance signal to background ratio, reduce the leading background by removing low $p_{T}>$ V candidates and jets with a high degree of separation by requiring dijet $p_{Tjj}>$~20GeV, $|\Delta\eta_{jj}|<1.5$. 

The main backgrounds are:
\begin{itemize}
\item $W+Jets$ ($\sigma$=3.1x104pb) - Dominant background, with a significant effort made to reduce its contribution.
\item $Z+Jets$ ($\sigma$=3.0x103pb) - Similar to $W+Jets$ but smaller in amount. One of the leptons doesn't pass the selection requirements.
\item $t\bar{t}$ ($\sigma$=163pb) - Two real W¡¯s and two b-jets. The contribution is reduced by anti-btagging. Alternatively, we can identify the hadronic W and use as a control sample to study the JES, JER.
\item Single Top ($\sigma$=85pb) - One (leptonic) W and a b-jet, with the contribution reduced by anti-btagging.
\item Multijet - The events are taken from the Data sideband with inverted isolation and the yield is estimated based on MET fit.
\end{itemize}
W vs Z mass difference is on the order of resolution of the detector, as a result the reconstructed signal is a combination $WW+WZ$. 
Reasonable agreement between data and MC is observed.

\section{Cross Section Measurement}
In order to extract the cross section we perform an unbinned maximum likelihood for the dijet mass $40 < m_{jj} < 150$~GeV. The shape templates taken from Monte Carlo (and multijet sideband), while the background yield contributions are free to float subject to Gaussian constraints described in Table.~\ref{tab:Table0}. Two separate fits for muon and electron event yields are performed and subsequently combined when evaluating the cross-section.
\begin{table}[bt]
\centering
\caption{Treatment of background dijet mass shapes and normalizations in a fit
to the data.
The cross section values are calculated with the programs cited on the
corresponding rows.
The background normalizations are constrained to Gaussian
distributions with the listed central values and widths. \label{tab:Table0}}
\begin{tabular} {lcl}
\hline
   Process             &    Shape      & Constraint on normalization \\
   \hline
   Diboson (WW+WZ)     &    MC  & Unconstrained \\
   W+Jets              &    MC  & $31.3nb \pm 5\%$ (NLO) \\
   $t\bar{t}$             &    MC  & $163pb \pm 7\%$ (NLO) \\
   Single top          &    MC  & $85pb \pm 5\%$(NNLL) \\
   Drell--Yan+jets     &    MC  & $3.05nb\pm 4.3\%$ (NNLO) \\
   Multijet (QCD)      &    data  & MET fit in data \\
   \hline
 \end{tabular}
\end{table}
The $W+Jets$ shape is a combination of: Default (MADGRAPH) MC, either Matrix Element - Parton Shower Matching Up ($\mu =2\mu_0$) or Matching Down ($\mu =0.5\mu_0$) MC, either Factorization Scale Up ($q' = 2 q_0$) or Scale Down ($q' =0.5 q_0$) MC. Namely, the $m_{jj}$ shape $\mathcal{F}_{W+Jets}$ is given by
\begin{equation}
F_{W+Jets} = \alpha\mathcal{F}_{W+Jets} (\mu_{0}^2, q'^2) + \beta\mathcal{F}_{W+Jets} (\mu'^2, q_{0}^2) + (1-\alpha-\beta)\mathcal{F}_{W+Jets} (\mu_{0}^2, q_{0}^2)\,,
\%
\end{equation}
where the parameters $\mu_0$ ($\mu'$) and $q_0$ ($q'$)
correspond to the default (alternative) values of $\mu$ and $q$,
respectively. The relative fractions $\alpha$ and $\beta$ are
free to vary during the fit and remain within the physical ranges ($0 \leq \alpha,\beta \leq 1$ and $1-\alpha-\beta\geq 0$). Likewise, the choice of Up or Down Sample is based on the best fit to the data with $\mu' = 2 \mu_0$ or $0.5 \mu_0$ ($q' = 2 q_0$ or $0.5 q_0$). Thus, we probe variations of a factor of two in both $\mu$ and $q$ (with the corresponding shape fluctuations accounted for when setting exclusion limits).

\begin{figure*}[tbh]
\includegraphics[width=0.30\textwidth]{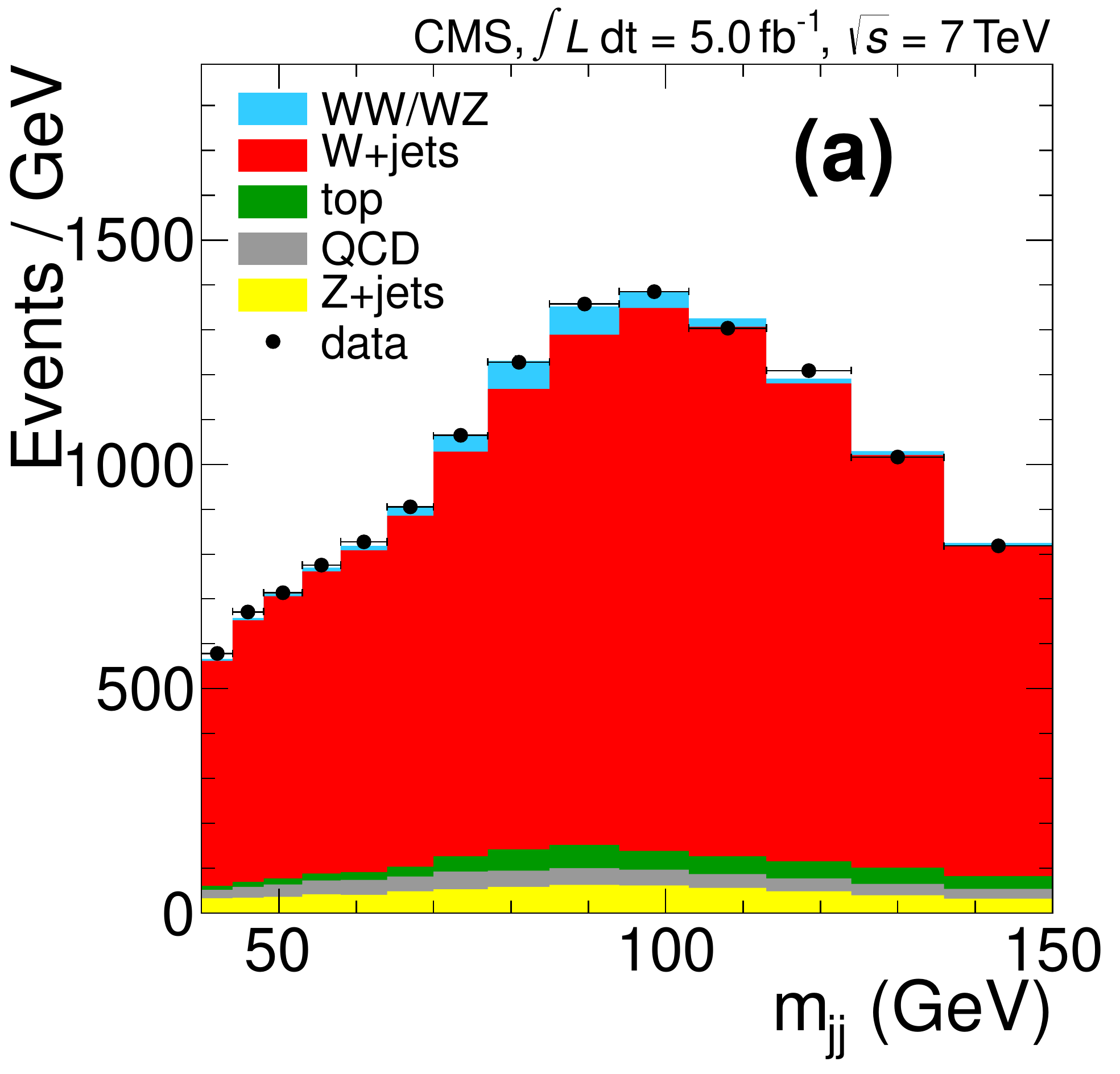}
\includegraphics[width=0.30\textwidth]{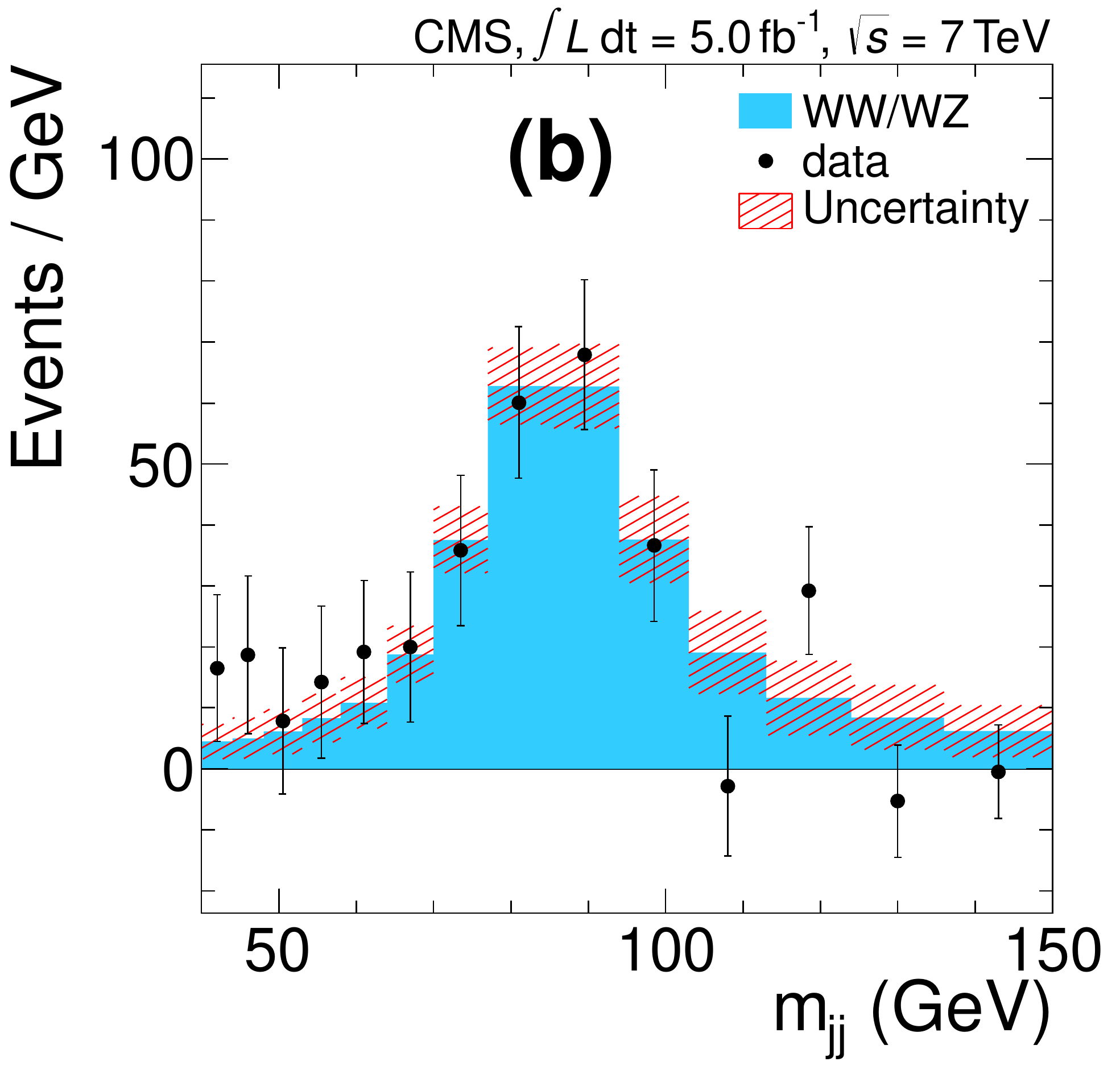}
\includegraphics[width=0.30\textwidth]{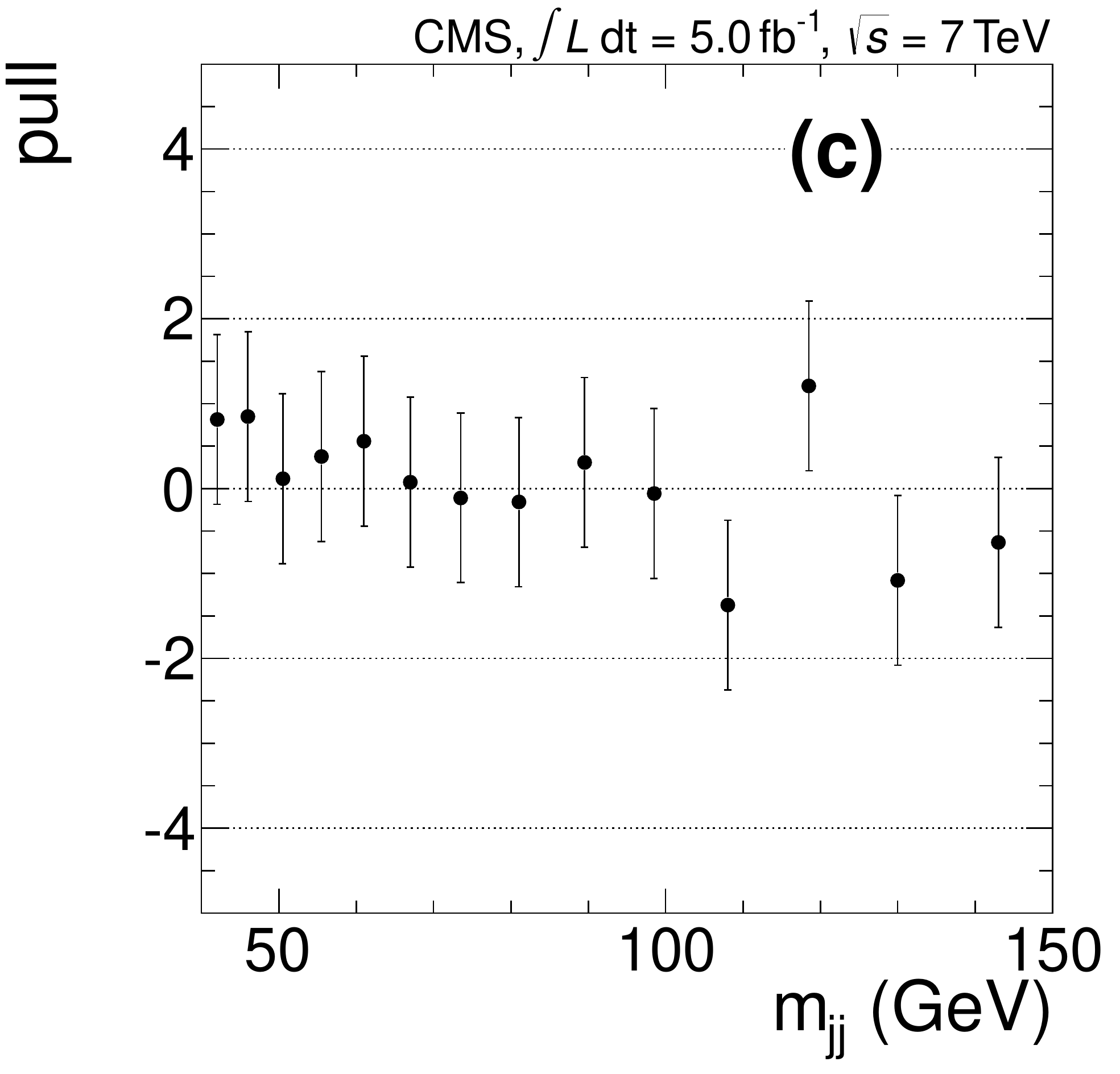}
\caption{(a) Distribution of the dijet invariant mass in data,
with the binning chosen based on the resolution and fit projections of the relevant components overlaid.
(b) The dijet invariant mass after subtraction of all components except the
electroweak WW+WZ processes.  The error bars represent the
statistical uncertainties and the hatched bands represent the systematic
uncertainties.  (c) The normalized
residual or pull: $(data - fit)/(fituncertainty)$.
}
\label{fig:Fig1}
\end{figure*}

The fit output is shown in Fig.~\ref{fig:Fig1} and summarized in Table.~\ref{tab:yields}. We extract $2682\pm 482$ $WW+WZ$ events out of $1.15\times 10^5$. Signal significance is $8.8\sigma$ using a simple likelihood ratio and $4.3\sigma$ using the profile likelihood ratio.
\begin{table*}[htbp]
  \caption{Event yields determined from a maximum-likelihood fit to the data.
  The total uncertainty is computed using the full covariance matrix. Owing to a higher kinematic threshold the
  product of acceptance $\times$ efficiency is smaller for the electron
  channel. The term $\mathcal{A}\varepsilon$ includes W and Z branching fractions.
  \label{tab:yields}}
\centering
\begin{tabular} {lll}
\hline
        Process          &    Muon channel                  & Electron channel \\
\hline
        Diboson (WW+WZ) &   1900 \,\,\,$\pm$ 370            & 800 \,\,\,\,\,\,$\pm$ 310   \\
        $W+Jets$     &   67380 $\pm$ 590		    & 31640 $\pm$ 850  \\
        $t\bar{t}$      &   1660 \,\,\,$\pm$ 120	    & 950 \,\,\,\,\,\,$\pm$ 70   \\
        Single top      &   650 \,\,\,\,\,\,$\pm$ 30        & 310 \,\,\,\,\,\,$\pm$  20   \\
        Drell--Yan+jets &   3610 \,\,\,$\pm$ 160	    & 1410 \,\,\,$\pm$ 60   \\
        Multijet  (QCD) &   300 \,\,\,\,\,\,$\pm$ 320       & 4190 \,\,\,$\pm$ 870   \\
\hline
    Data                &    75419           &  39365           \\
    Fit $\chi^2/N_{dof}$ (probability) &   9.73/12 (0.64)     &  5.30/12 (0.95)  \\
\hline
    Acceptance $\times$ efficiency ($\mathcal{A}\varepsilon$) & $(5.15 \pm 0.24) \times 10^{-3}$ & $(2.63 \pm 0.12) \times 10^{-3}$ \\
\hline
    Expected WW+WZ yield from simulation & 1700 $\pm$ 60 & 870 $\pm$ 30  \\
\hline
  \end{tabular}
\end{table*}

We validate the fitter by performing pseudo-experiments (with correlations taken into account)
and correct the yields and errors based on the resulting pull distributions. The procedure also accounts for
the uncertainty due to limited MC. $W+Jets$shape error, as well as uncertainties due to the choice of ME-PS matching and Factorization/Renormalization scale are covered by the empirical model. The systematics due to trigger efficiency (1\%), lepton reconstruction and selection efficiency (2\%), jet energy scale (0.6\%), missing transverse energy resolution (0.5\%), fit uncertainty (0.2\%), luminosity determination (2.2\%) as well as theory uncertainty on acceptance (4\%) are included. We measure the $WW+WZ$ cross section to be $68.9 \pm 8.7$(stat) $\pm 9.7$(syst) $\pm 1.5$(lumi)~pb, consistent with the SM prediction of $65.6\pm 2.2$pb.

\section{aTGC Limits}
\begin{figure*}[tbh]
  {\centering
    \includegraphics[width=0.48\textwidth]{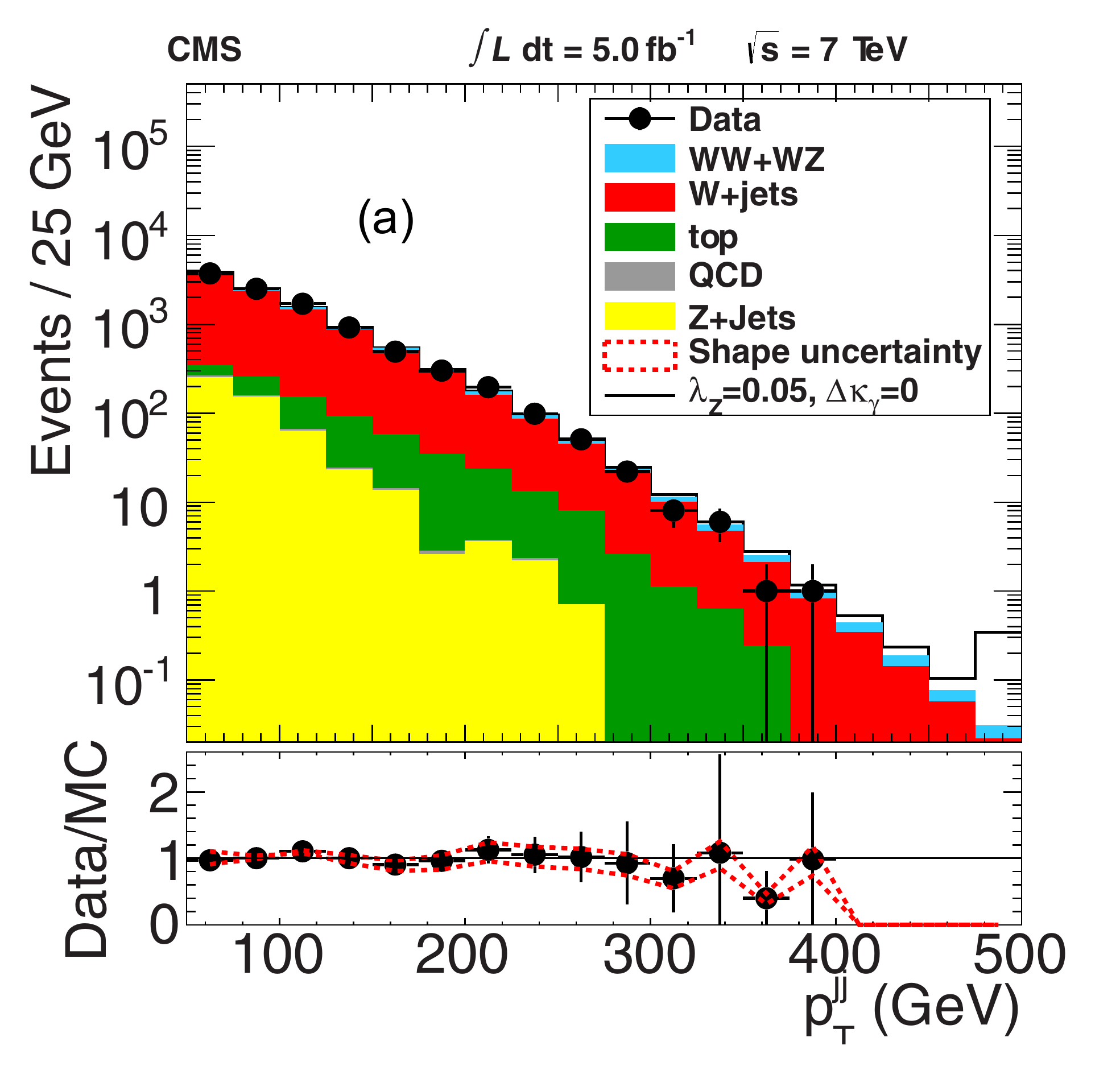}
    \includegraphics[width=0.48\textwidth]{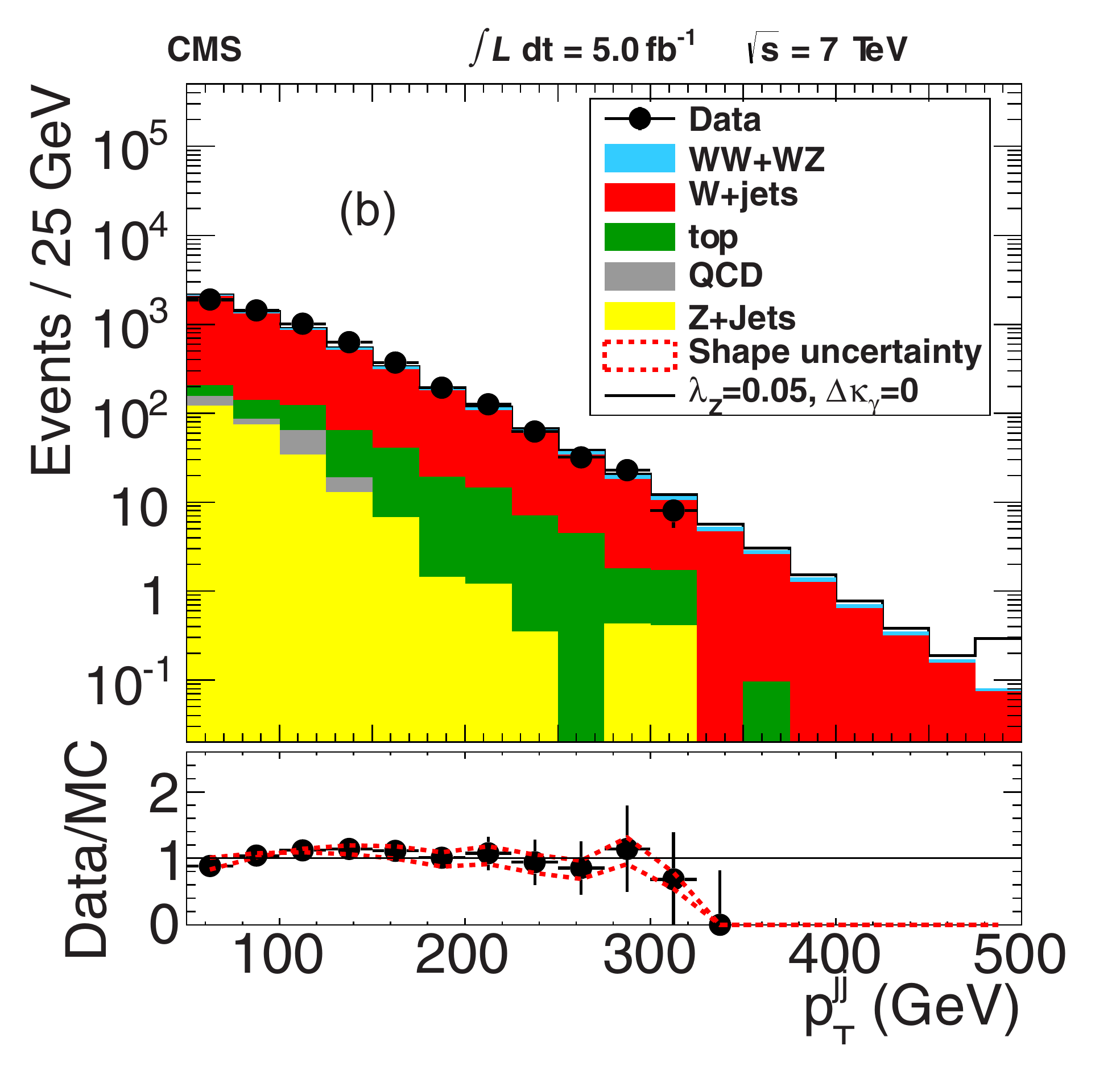}
    \caption{Dijet $p_T$ distributions for (a) the muon and (b) the electron
    channels after full selection and with the requirement
    $75 < m_{jj} < 95$~GeV.
    The stacked histogram shapes are taken from simulation
    or, where applicable, from data-driven estimates.
    They are normalized according to the fit to the observed
    $m_{jj}$ spectrum in data. Below we show the Data/MC ratio with the (dashed) red lines corresponding to the shape uncertainty.
    The last bin includes the overflow.
    }
    \label{fig:Fig2}}
\end{figure*}

\begin{figure}[tbh]
  {\centering
    \includegraphics[width=0.48\textwidth]{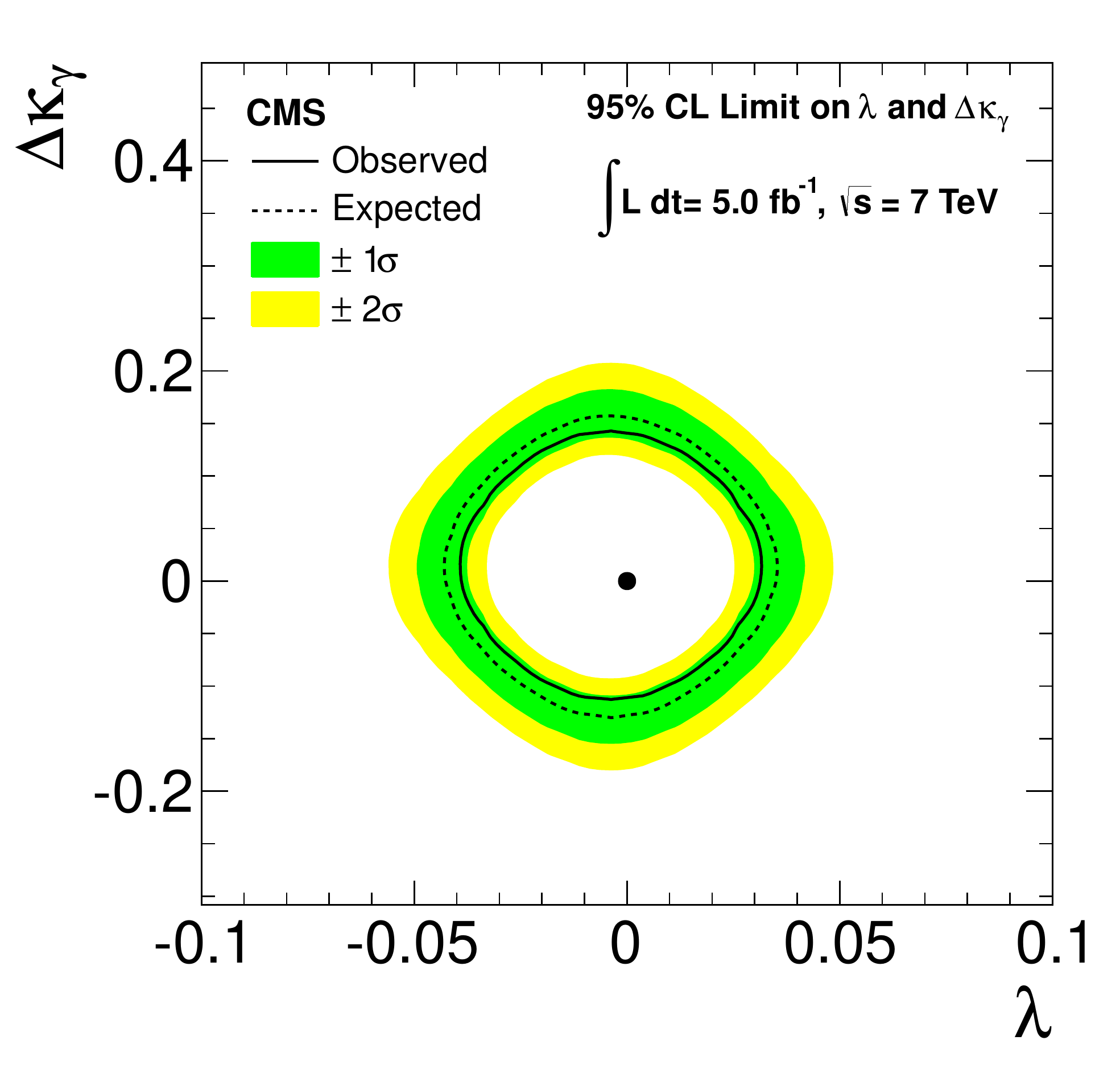}
    \caption{Observed (solid) and expected (dashed) exclusion limits at 95\% CL for
    anomalous triple gauge couplings.
    The dark green and light yellow bands correspond to
    the one and two sigma intervals, respectively, in the
    expected limit distribution.
    The SM expectation is shown by the solid dot.
    }
    \label{fig:Fig3}}
\end{figure}

 In the standard model the the triple-gauge couplings 
$\gamma{WW}$ and $ZWW$, responsible for $s$-channel production of $WW$ events, are defined up to 
an overall constant. Corresponding new physics processes would naturally affect the measured $WW$ cross section and can be parametrized at low energy by an effective 
Lagrangian containing a subset of the 14 possible terms consistent with Lorentz invariance~\cite{Hagiwara1987253}:
\begin{equation}
   \begin{array}{ccl}
    \frac{{\mathcal L}_{eff}^{VWW}}{g_{VWW}} & = & i g_{1}^{V} (W_{\mu\nu}^{*}W^{\mu}V^{\nu} - 
     W_{\mu}^{*}V_{\nu}W^{\mu\nu})+i{\kappa}_{V}W_{\mu}^{*}W_{\nu}V^{\mu\nu} + 
     i\frac{\lambda_{V}}{M_{W}^{2}} W_{\lambda,\mu}^{*}W_{\nu}^{\mu}V^{\nu\lambda} \\ & - 
     & g_{4}^{V}W_{\mu}^{*}W_{\nu}(\partial^{\mu}V^{\nu} + \partial^{\nu}V^{\mu}) + 
     g_{5}^{V}\epsilon^{\mu\nu\lambda\rho}(W_{\mu}^{*}\partial_{\lambda}W_{\nu} -
     \partial_{\lambda}W^{*}_{\mu}W_{\nu})V_{\rho} \\ & + 
     & i\tilde{\kappa}_{V}W^{*}_{\mu}W_{\nu}\tilde{V}^{\mu\nu} + 
     i\frac{\tilde{\lambda}_{V}}{M_{W}^{2}}W^{*}_{\lambda\mu}W^{\mu}_{\nu}\tilde{V}^{\nu\lambda},
   \label{eq:effLang}
   \end{array}{}
\end{equation}
where $\epsilon_{\mu\nu\lambda\rho}$ is the fully antisymmetric $\epsilon$ - tensor, 
$W$ denotes the $W$ boson field, $V$ denotes the photon or $Z$ boson field, 
$V_{\mu\nu}=\partial_{\mu}V_{\nu}-\partial_{\nu}V_{\mu}$, 
$W_{\mu\nu}=\partial_{\mu}W_{\nu}-\partial_{\nu}W_{\mu}$, 
$\tilde{V}_{\mu\nu}=1/2(\epsilon_{\mu\nu\lambda\rho}V^{\lambda\rho})$, 
$g_{\gamma{WW}}=-e$ and $g_{ZWW}=-e\cot\theta_{w}$. The equation is further simplified in the presence of following symmetries:
\begin{enumerate}
\item C and P invariance: $g_{5}^{V}=g_{4}^{V}=\tilde{\kappa}_{V}=\tilde{\lambda}_{V}=0$
\item Electro-magnetic gauge invariance: $g_{1}^{\gamma}=1$.
\item $SU(2)\times U(1)$ gauge invariance in the presence of a Higgs doublet and considering up to dimension 6 operators: 
  \begin{equation}
    \begin{array}{ccl}
      \Delta\kappa_{Z} = \Delta g^{Z}_{1}-\Delta\kappa_{\gamma}\cdot tan^{2}\theta_{w} & and &  
      \lambda_{Z} = \lambda_{\gamma} = \lambda, 
      \label{eq:lep}
    \end{array}{}
  \end{equation}
  where $\theta_{w}$ is the weak-mixing angle and $\Delta$ denotes deviation from the standard model.
\end{enumerate}.
Furthermore, $\Delta g_{1}^{Z}$ is small and we take it to be 0. The limits on the remaining two parameters are then set by examining the hadronic $V\to jj$ $p_T$ distributions. We place a $75 < m_{jj} < 95$~GeV cut to enhance signal purity and normalize the backgrounds based on fit results. The distributions are shown in Fig.~\ref{fig:Fig2}, where the last bin includes overflow events. We model aTGC dependence by reweighting SM Diboson MC to MCFM predictions;account for systematics due to luminosity, signal selection efficiency, signal shape, from the normalization and shape of the SM processes; and use a CLs (a modified frequentist construction) with profile likelihood as a test statistic is used to set limits (Fig.~\ref{fig:Fig3}).


\section{Conclusions}

We measure of the sum of the inclusive WW and WZ production cross sections
by using events containing a leptonically decaying W and two jets. This is the first measurement of WW+WZ production in pp collisions using
this signature. Its measured to be $\sigma(WW+WZ) = 68.9 \pm 8.7$(stat) $\pm 9.7$(syst) $\pm 1.5$(lumi)~pb,
consistent with the SM prediction.  No evidence for anomalous triple gauge couplings is found, and we set the 1-Dimensional limits of $ -0.038 < \lambda < 0.030$,
$ -0.11 < \Delta{\kappa_\gamma} < 0.14$ at 95\% C.L. The result is an improvement upon the leptonic
channels and competitive with the LEP combination~\cite{aTGCtwiki}.

\Acknowledgments

\end{document}

%% file: econfmacros.tex
\def\beq{\begin{equation}}
\def\eeq#1{\label{#1}\end{equation}}
\def\eeqn{\end{equation}}

\def\beqa{\begin{eqnarray}}
\def\eeqa#1{\label{#1}\end{eqnarray}}
\def\eeqan{\end{eqnarray}}

\let\bar=\overbar

\def\Dslash{\not{\hbox{\kern-4pt $D$}}}
\def\dslash{\not{\hbox{\kern-2pt $\del$}}}

\def\msb{{\bar{\ssstyle M \kern -1pt S}}}

